\PassOptionsToPackage{prologue,dvipsnames}{xcolor}
\documentclass[sigconf, screen]{acmart}

\usepackage{graphicx}  %
\usepackage{multirow}
\usepackage{algorithm,algpseudocode}
\usepackage{subcaption}
 
\usepackage{amssymb}
\usepackage{enumitem}
\usepackage{float}
\usepackage{diagbox}
\usepackage{tikz}
\usepackage[normalem]{ulem}
\useunder{\uline}{\ul}{}
\usepackage{float}
\usepackage{hyperref}

\AtBeginDocument{%
  \providecommand\BibTeX{{%
    \normalfont B\kern-0.5em{\scshape i\kern-0.25em b}\kern-0.8em\TeX}}}

\copyrightyear{2022} 
\acmYear{2022} 
\setcopyright{acmcopyright}\acmConference[MM '22]{Proceedings of the 30th ACM
International Conference on Multimedia}{October 10--14, 2022}{Lisboa, Portugal}
\acmBooktitle{Proceedings of the 30th ACM International Conference on Multimedia
(MM '22), October 10--14, 2022, Lisboa, Portugal}
\acmPrice{15.00}
\acmDOI{10.1145/3503161.3548368}
\acmISBN{978-1-4503-9203-7/22/10}

\acmSubmissionID{mmfp2843}

\begin{document}
\title{SongDriver: Real-time Music Accompaniment Generation \\ without Logical Latency nor Exposure Bias}

\author{Zihao Wang}
\authornote{Both authors contributed equally to this research and share the co-first authorship.}
\affiliation{%
  \institution{Zhejiang University}
   \city{Hangzhou}
   \country{China}
}
\email{carlwang@zju.edu.cn}

\author{Qihao Liang}
\authornotemark[1]
\affiliation{%
  \institution{Zhejiang University}
   \city{Hangzhou}
   \country{China}
}
\email{qhliang@zju.edu.cn}

\author{Kejun Zhang}
\authornote{Corresponding author.}
\affiliation{%
  \institution{Zhejiang University}
  \institution{Alibaba-Zhejiang University Joint Institute of Frontier Technologies}
   \city{Hangzhou}
   \country{China}
}
\email{zhangkejun@zju.edu.cn}

\author{Yuxing Wang}
\affiliation{%
  \institution{Zhejiang University}
   \city{Hangzhou}
   \country{China}
}
\email{3180101871@zju.edu.cn}

\author{Chen Zhang}
\affiliation{%
  \institution{Zhejiang University}
   \city{Hangzhou}
   \country{China}
}
\email{zc99@zju.edu.cn}

\author{Pengfei Yu}
\affiliation{%
  \institution{Jingchu University of Technology}
   \city{Hubei}
   \country{China}
}
\email{cgoxopx@qq.com}

\author{Yongsheng Feng}
\affiliation{%
  \institution{Shandong University}
   \city{Weihai}
   \country{China}
}
\email{201900800360@mail.sdu.edu.cn}

\author{Wenbo Liu}
\affiliation{%
  \institution{Zhejiang University}
   \city{Hangzhou}
   \country{China}
}
\email{3190102475@zju.edu.cn}

\author{Yikai Wang}
\affiliation{%
  \institution{Zhejiang University}
   \city{Hangzhou}
   \country{China}
}
\email{wangyik@zju.edu.cn}

\author{Yuntai Bao}
\affiliation{%
  \institution{Zhejiang University}
   \city{Hangzhou}
   \country{China}
}
\email{3190105994@zju.edu.cn}

\author{Yiheng Yang}
\affiliation{%
  \institution{Zhejiang University}
   \city{Hangzhou}
   \country{China}
}
\email{yvonneyoung@zju.edu.cn}





\renewcommand{\shortauthors}{Zihao Wang et al.}





\begin{abstract}
Real-time music accompaniment generation has a wide range of applications in the music industry, such as music education and live performances. However, automatic real-time music accompaniment generation is still understudied and often faces a trade-off between logical latency and exposure bias. In this paper, we propose SongDriver, a real-time music accompaniment generation system without logical latency nor exposure bias. Specifically, SongDriver divides one accompaniment generation task into two phases: 1) The arrangement phase, where a Transformer model first arranges chords for input melodies in real-time, and caches the chords for the next phase instead of playing them out. 2) The prediction phase, where a CRF model generates playable multi-track accompaniments for the coming melodies based on previously cached chords. With this two-phase strategy, SongDriver directly generates the accompaniment for the upcoming melody, achieving zero logical latency. Furthermore, when predicting chords for a timestep, SongDriver refers to the cached chords from the first phase rather than its previous predictions, which avoids the exposure bias problem. Since the input length is often constrained under real-time conditions, another potential problem is the loss of long-term sequential information. To make up for this disadvantage, we extract four musical features from a long-term music piece before the current time step as global information. In the experiment, we train SongDriver on some open-source datasets and an original àiSong Dataset built from Chinese-style modern pop music scores. The results show that SongDriver outperforms existing SOTA (state-of-the-art) models on both objective and subjective metrics, meanwhile significantly reducing the physical latency.
\end{abstract}



\maketitle

\section{INTRODUCTION}

Real-time accompaniment improvisation is an intricate task with high entry barriers even for human musicians. To make accompaniment improvisation easier and accessible to not merely professionals, but also amateur music lovers, automatic real-time music accompaniment generation models are worth further researching.

One of the central problems in automatic real-time accompaniment generation tasks arises from the compromise between logical latency \cite{havers2019driven} and exposure bias \cite{liu2018generative}. The logical latency is the time difference between accompaniment generation and playback. The exposure bias in real-time music generation tasks, refers to the quality decrease due to error accumulation: Sequence models often rely on their previous predictions when making a new inference. Therefore, an improper prediction may continuously influences its following timesteps, which is the said error accumulation. 

Previous automatic real-time accompaniment models are generally one-phase, and can be divided into two categories on the basis of the said two problems: 1) Latency models \cite{verma2012real}\cite{wallace2019comparing}, which avoid the exposure bias but suffer from the logical latency by arranging the accompaniment each time the input reaches a preset length. 2) Bias models \cite{cunha1999intelligent}\cite{jiang2020rl}, which eliminate the logical latency but face the exposure bias by generating accompaniments for the upcoming melody.

\begin{figure}[h]
  \centering
  \vspace{-0.3cm}
  \includegraphics[width=0.7\linewidth]{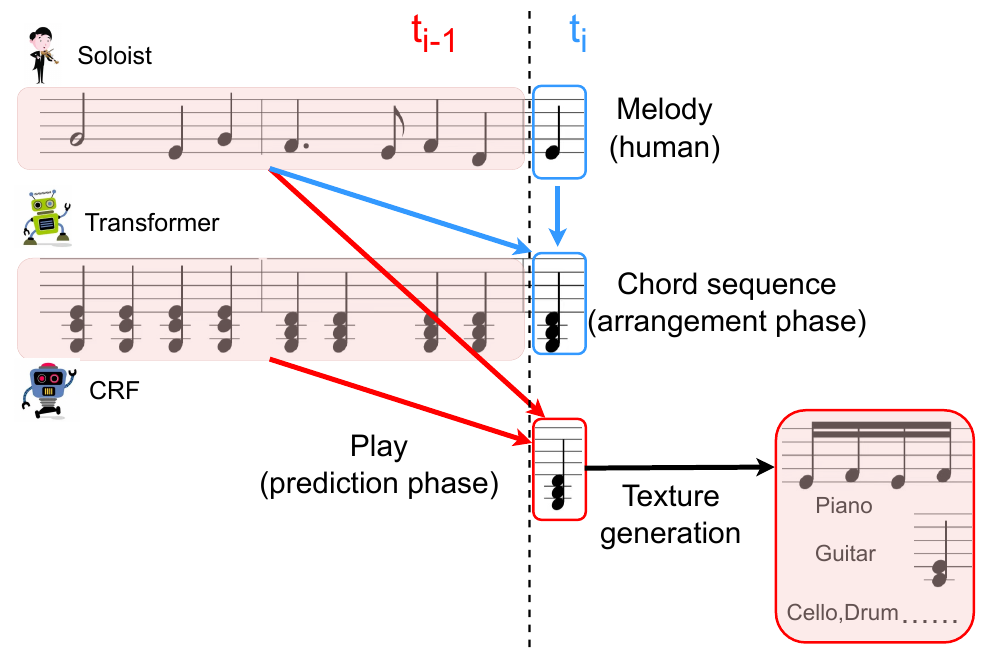}
  \vspace{-0.4cm}
  \caption{An overview of the mechanism of SongDriver. The multi-track accompaniment in the red rectangle regions is generated at time step i-1, while the chords in the blue rectangle regions are generated at time step i.}
  \label{Intro}
    \vspace{-0.3cm}
\end{figure}

In this paper, SongDriver divides the generation process into two successive phases: 1) the arrangement phase and 2) the prediction phase. The arrangement phase employs a Transformer \cite{vaswani2017attention} model: The model reads instreaming melody inputs and correspondingly arranges chords for the former beats. The chords generated in this phase will be cached in a sequence rather than being played. The prediction phase uses a CRF \cite{lafferty2001conditional} model to predict playable chords that are most likely to harmonize with the upcoming melody, with the reference to the cached chords as well as the previous melody. These predicted chords are directly played instead of being cached. To enhance the musicality of the generated music, we also design some texture generation patterns to transform the predicted chords into multi-track pieces of accompaniment. In Fig.\ref{Intro}, we provide an overview of the SongDriver’s mechanism.

With this two-phase strategy, SongDriver solves the logical latency and exposure bias problems at the same time: The accompaniment generated by SongDriver is specifically for the upcoming melody, and is played exactly when the corresponding melody starts. This eliminates the logical latency and improves the immediacy of accompaniment generation. Besides, the prediction phase keeps referring to the cached chords from the arrangement phase rather than its previous predictions, which avoids exposure bias and thus can fully improve musicality. Furthermore, this mechanism also highly improves model performance by incorporating the advantages of both models while eliminating their potential problems.

In real-time music accompaniment generation, the length of the melody sequence increases with time. To reduce the difficulty of modeling long-term sequences, we need to constrain the length of input melody, which leads to a loss of long-term musical information. To compensate for this drawback, we extract four musical features from a long-term music piece ahead of the current time step to serve as global information. 1) the \textbf{weighted notes: }the notes that are more important in the melody. 2) the \textbf{weighted factors}: the overall features of a melody piece. 3) the \textbf{terminal chords: }the chords marking the ends of musical phrases. 4) the \textbf{structural chords: }the chords which maintain the stability and continuity of the musical accompaniment.

Apart from the open-source musical datasets, we also build an original Chinese-style àiSong Dataset from scratch, which is mainly based on the Chinese folk pentatonic tuning. 

We train SongDriver on both open-source and àiSong Datasets, and evaluate the result both objectively and subjectively. The results demonstrate that SongDriver can respond to the melody in real time input and generate music accompaniment with imperceptible logical latency and relatively high musicality.

In summary, our main contributions are as follows:

1.\, We propose a real-time music accompaniment generation mechanism that simultaneously eliminates logical latency and exposure bias.

2.\, We introduce four musical features and their extraction methods, making up for the loss of long-term music structure information under real-time conditions.

3.\, We build an original àiSong Dataset from scratch based on Chinese-style modern pop music scores.

4.\, We highly improve the model performance thanks to the parallel two-phase strategy, where the physical latency infinitely approaches the processing time of the second-phase model (CRF).

5.\, We also distill traditional Transformer to be suitable for real-time tasks, which helps us maintain high generation quality while keeping the extremely low physical latency of CRF.

\section{RELATED WORK}

Automatic music generation has been a heated research topic for more than half a century \cite{hiller1979experimental}. In recent years, the rapid development of neural networks has resulted in more and more deep learning-based music generation models, including CNN \cite{yang2017midinet}, RNN \cite{simon2017performance}, and GAN \cite{dong2018musegan}. Transformer, based solely on self-attention mechanisms, has been proved to be the state-of-the-art model for language generation by various evaluations \cite{huang2018music}\cite{donahue2019lakhnes}\cite{dai2019transformer}. Some research successfully employed the Transformer to compose music pieces, such as Music Transformer \cite{huang2018music} and MuseNet \cite{payne2019musenet}. Music accompaniment generation is an important topic of the automatic music generation, and can be categorized into two groups: non-real-time accompaniment generation and real-time accompaniment generation.

\textbf{Non-real-time Automatic Music Accompaniment Generation. }Many previous researches have proposed a series of effective systems for non-real-time music accompaniment generation. Early studies in this field only select the pre-recorded audio tracks to accompany human soloists  \cite{dannenberg1984line}\cite{raphael2010music}. Later, the automatic accompaniment generation was implemented with HMM model, such as MySong\cite{simon2008mysong}. Some recent studies resort to deep learning methods to generate accompaniment for the input melody. MuseGAN\cite{dong2018musegan} employs GAN-based model to generate multi-track music, which can also be extended as a human-AI cooperative system, but the training of GAN-based model is difficult and often requires big data. PopMAG\cite{ren2020popmag} uses Transformer to capture the long-term features of music sequences, generating multi-instrument pop music accompaniment.

\begin{figure}[htbp]
  \centering
  \vspace{-0.4cm}
  \includegraphics[width=0.70\linewidth]{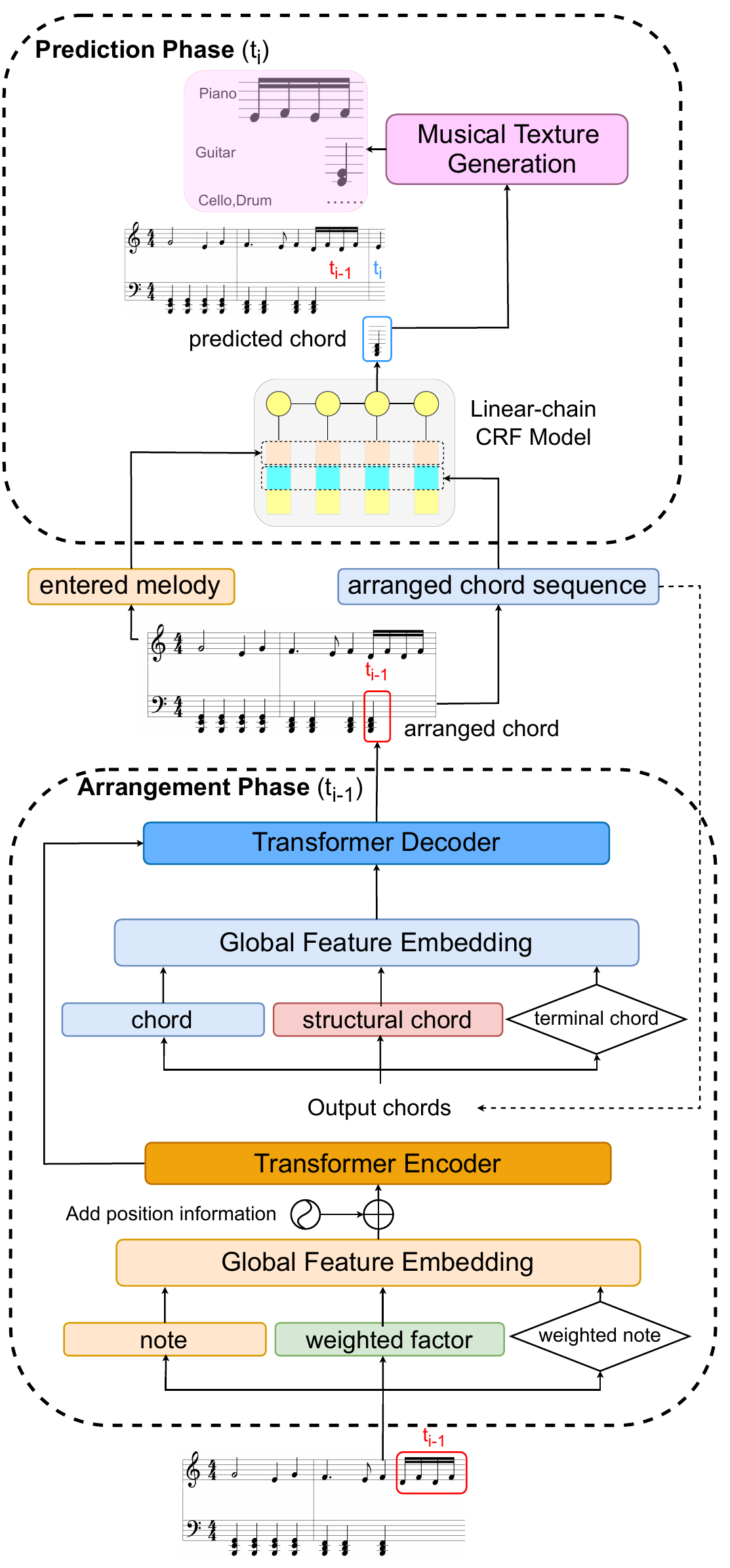}
  \vspace{-0.4cm}
  \caption{The overall architecture of SongDriver. Only the multi-track accompaniment (music in the pink zone at the top) after the texture generation will be played.}
  \vspace{-0.3cm}
  \label{Overall}
\end{figure}

\textbf{Real-time Automatic Music Accompaniment Generation.}
In contrast with non-real-time tasks, only a few explorations into real-time automatic music accompaniment generation can be found. These existing models could be classified into two categories: latency models and bias models. Latency models can be represented by \cite{wallace2019comparing}, which uses HMM and RNN to ensure the accuracy of the improvised accompaniment. But the accompaniment often slightly lags behind its corresponding melody, thus leading to a logical latency. The bias models, on the other hand, usually adopt forecasting strategies. For example, RL-Duet \cite{jiang2020rl} uses deep reinforcement learning to predict the next machine note based on the previous human and machine music parts. Another example\cite{Lin2021Markov} simplifies HMM into Markov Chains to support real-time music accompaniment generation, using previous melodies and predicted results to infer the chords of the next time step. These bias models eliminate the logical latency by predicting the accompaniment for the subsequent melody, but their performance may decline because of the exposure bias. Therefore, our goal in designing SongDriver is to tackle both the logical latency and exposure bias, making up for the limitations of existing models.

\textbf{Additional Information Guidance in Music Generation. } Some previous works employ unconditional autoregressive models, which are majorly trained with only sequences of note tokens, and thus cannot well capture the high-level musical features crucial to musical quality. To tackle this problem, some scholars intend to generate music by introducing additional conditions. For example, MuseNet \cite{payne2019musenet} uses a conditional scenario to guide Transformer to generate musical compositions with patterns of harmony, rhythm and style information. Besides, MuseMorphose\cite{wu2021musemorphose} proposes three mechanisms to constrain Transformer decoder, where the pre-attention conditioning is similar to our guiding strategy, under which the conditions enter the Transformer decoder before all the self-attention layers. These works all demonstrate the feasibility of using additional conditions in music generation. Therefore, we extracted four types of musical features as global information to guide the model to capture crucial musical features.

\section{METHOD}

Fig.\ref{Overall} shows the two-phase architecture of SongDriver. The arrangement phase includes a Transformer model with four musical features embedded. The prediction phase consists of a CRF architecture and also texture generation algorithms that output playable accompaniments of higher musicality. 

\subsection{Arrangement Phase}

\subsubsection{Feature Embedding}

\begin{figure}[htbp]
  \centering
  \includegraphics[width=0.80\linewidth]{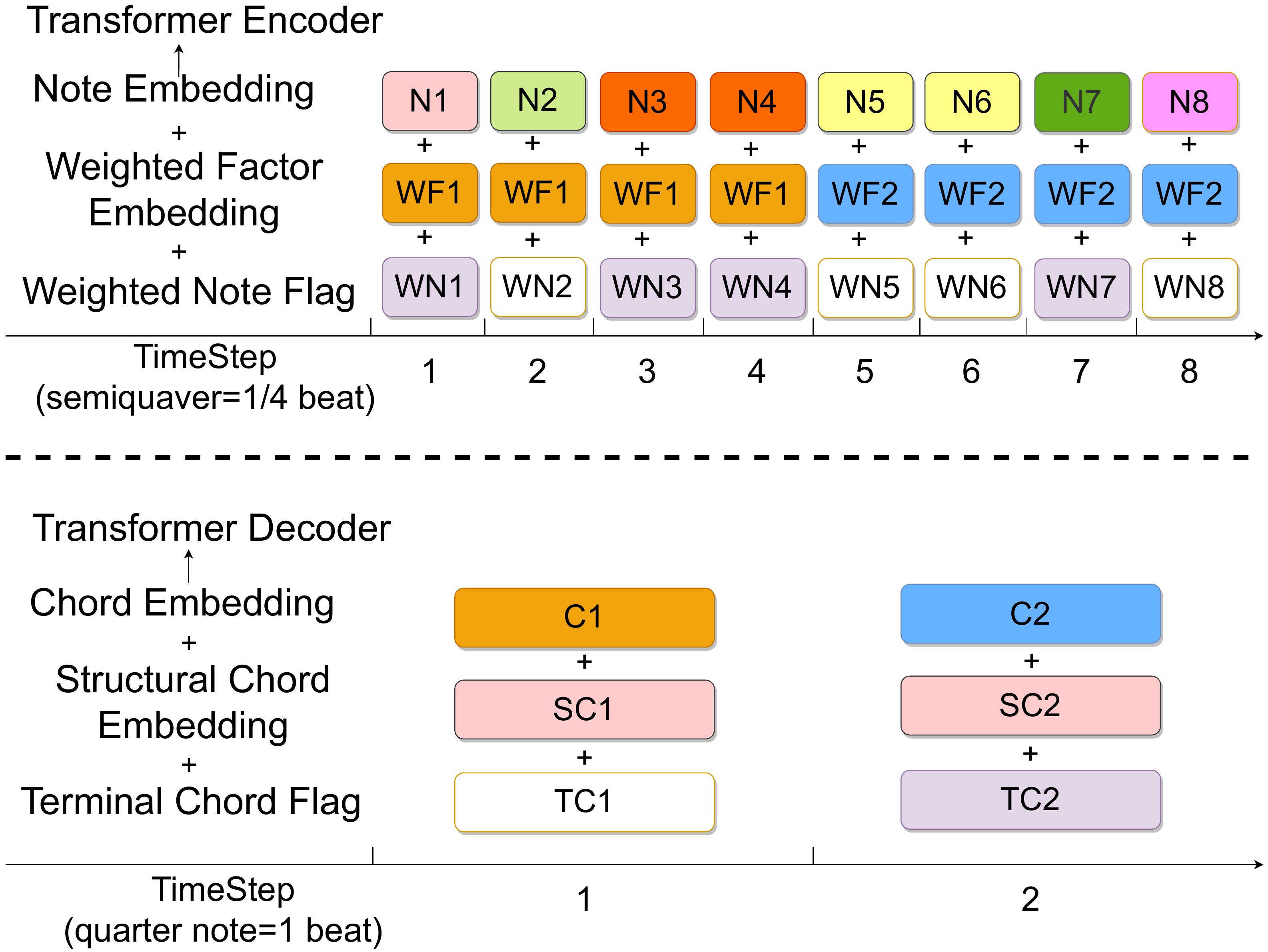}
  \vspace{-0.5cm}
  \caption{The embedding strategy of the musical features. The subscript is incremented by one for each extraction. And the same color indicates the same value.}
  \label{Embedding}
  \vspace{-0.4cm}
\end{figure}

Inspired by Compound Word \cite{hsiao2021compound}, we apply the extracted four features as global information in the arrangement phase. This can compensate for the loss of the long-term sequence information and guide the Transformer to capture the important musical features. We adopt a strategy similar to pre-attention \cite{wu2021musemorphose}, where the conditions enter the Transformer encoder or decoder before all the self-attention layers: Weighted factors and weighted notes are embedded with each sixteenth note before being input into the \textbf{Transformer Encoder }, while structural chords and terminal chords are embedded with each arranged chord as \textbf{Transformer Decoder }inputs. The embedding at each timestep is shown in Fig.\ref{Embedding}.

\subsubsection{Chord Arrangement}

Thanks to the self-attention mechanism, Transformer is able to capture long-term dependencies in sequential data. Therefore, we employ Transformer in this phase for sake of a higher musicality. To reduce the difficulty of modeling long sequences in real-time, the length of the input melody stream is constrained, and here we specify the length to be one beat. Although Transformer cannot see the complete melody sequence under this condition, the proposed musical features can provide contextual information and guide Transformer to capture important musical structures. We also modify the wait-k \cite{ren2020simulspeech} strategy used in the real-time translation task to ensure the input having sufficient information: Instead of arranging once per sample, Transformer will wait for 4 samples, that is, arranging once per beat.

The output of the arrangement phase at each time step is a chord arranged for the former beat of melody. This chord will not be played and will be cached into a chord sequence as an input of the decoder for the next generation. The chord sequence will be used as references in the predicted phase, as shown in the middle part of Fig. \ref{Overall}.

\subsection{Prediction Phase}

\vspace{-0.1cm}
\begin{figure}[htbp]
  \centering
  \includegraphics[width=0.9\linewidth]{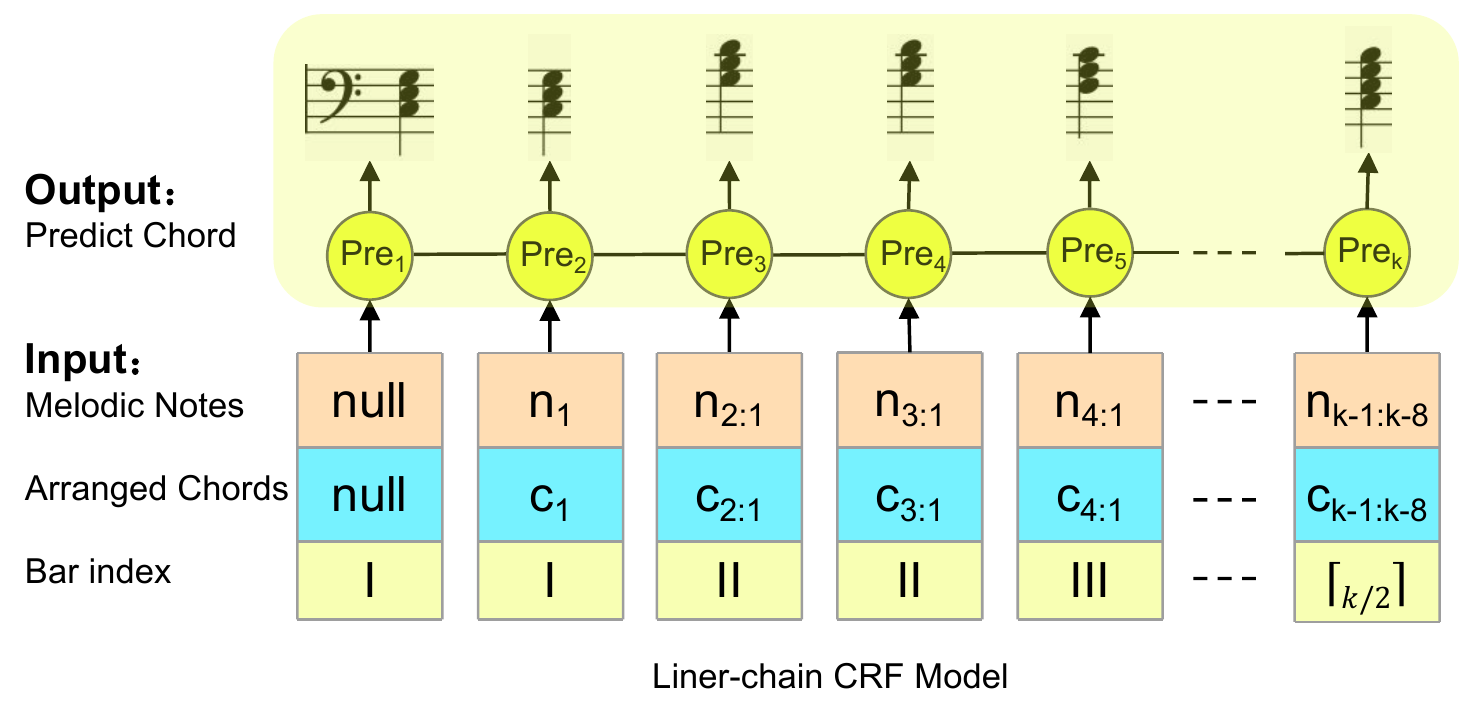}
  \vspace{-0.5cm}
  \caption{
  The architecture of the CRF model in the prediction phase. $n_k$ is the first longest melodic note on time step k. $c_k$ is the arranged chord on time step k. Bar index indicates the number of bars. $Pre_k$ is the predicted chord on time step k.}
  \label{CRF}
  \vspace{-0.5cm}
\end{figure}

\subsubsection{Chord Prediction}

In the prediction phase, we firstly use Conditional Random Fields (CRF) to predict the chord for the upcoming melody. We adopt the linear-chain form of CRF, which is more suitable for sequential data. Based on the cached chord sequence from the arrangement phase and also the previous melody, CRF calculates the conditional probability to predict the next chord. The architecture of the predicted phase is shown in Fig. \ref{CRF}. Regarding the input at each time step, we take the music information from the previous eight beats, including the beat’s bar index, arranged chord for the beat, and the longest note in this beat. The chords obtained in the prediction phase are only played instead of being used as references for the subsequent predictions, which avoids the exposure bias. 

\subsubsection{Texture Generation}
\begin{figure}[htbp]
  \centering
  \vspace{-0.2cm}
  \includegraphics[width=\linewidth]{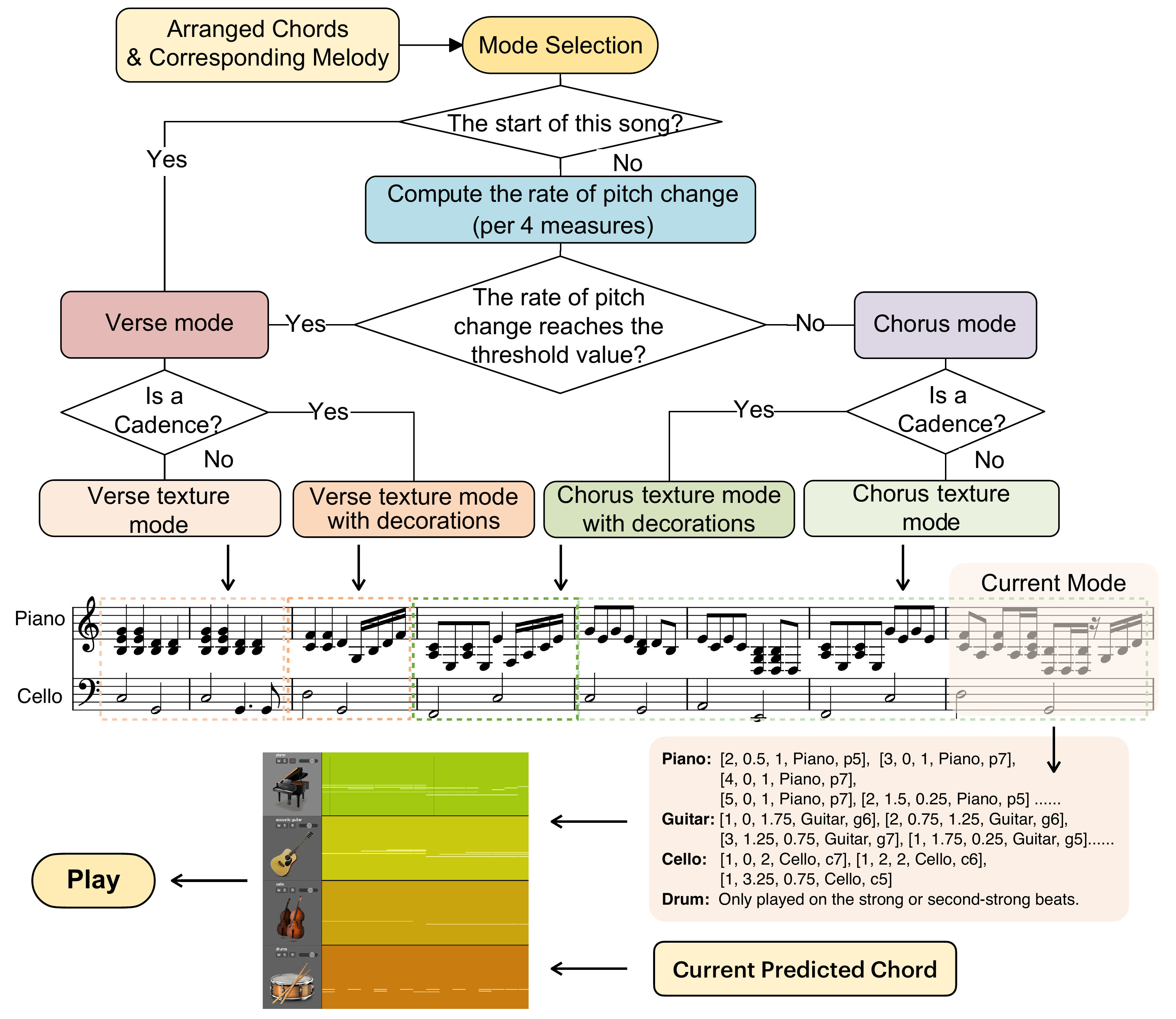}
  \vspace{-0.6cm}
  \caption{The process of texture generation}
  \label{Texture}
  \vspace{-0.5cm}
\end{figure}

Texture arrangement can enhance the expressiveness of harmonics, which is also a basic technique broadly utilized among music professionals. Therefore, we design various texture patterns which may change according to the melodic progression and musical phrase-level structures. 
The choice of texture pattern and its application to the current chord are briefly shown in Fig.\ref{Texture}.

Different texture generation patterns correspond to different parts of music. We choose the most appropriate texture pattern for different musical structures, transforming the predicted chords into multi-track accompaniments for playback. For every four bars, we determine whether the current time step is in the verse or in the chorus parts by the variation rate of pitches. Next, we utilize the terminal chord to segment the music at the phrase level to switch between multiple textures. Within a phrase, regular textures will be played in a pre-set order. And between different phrases, a bar will be played in a decorative texture. The details of texture patterns are included in Section \ref{section:A} of Appendices.

\section{THE EXTRACTION OF MUSICAL FEATURES}

The proposed four musical features are based on Schenkerian analysis method and the fundamentals of music theory. The features provide SongDriver with global music information to compensate for the loss of the long-term dependencies. Detailed explanations regarding all the following equations, illustrations and algorithms in this section are listed in the Section \ref{section:B} of Appendices.

\subsection{Weighted Note}

\begin{equation}
T = N_{strong} \cup  N_{next-strong }
\label{eq:T}
\end{equation}
\begin{equation}
C = {N_{weak}} \cap( L(N)>\frac{3}{2}steps )
\label{eq:C}
\end{equation}
\begin{equation}
L = {N_{latest}} \cap N_{longest}
\label{eq:L}
\end{equation}
\begin{equation}
Weighted\,\, Note = (T\cap \overline C) \cup ({\overline T} \cap C \cap L)
\label{eq:WN}
\end{equation}

In the melody, the importance of notes is not equal. Some notes are decisive to the musicality, and we define these notes as the weighted notes. Weighted notes shown in Equation (\ref{eq:WN}) are derived from three musical concepts, accent(\ref{eq:T}), syncopation(\ref{eq:C}) and long note(\ref{eq:L}). During the generation process, the weighted notes can mark the importance degree of the current note and optimize the model’s attention distribution.

\subsection{Weighted Factor}

\begin{figure}[H]
  \centering
  \vspace{-0.4cm}
  \includegraphics[width=0.7\linewidth]{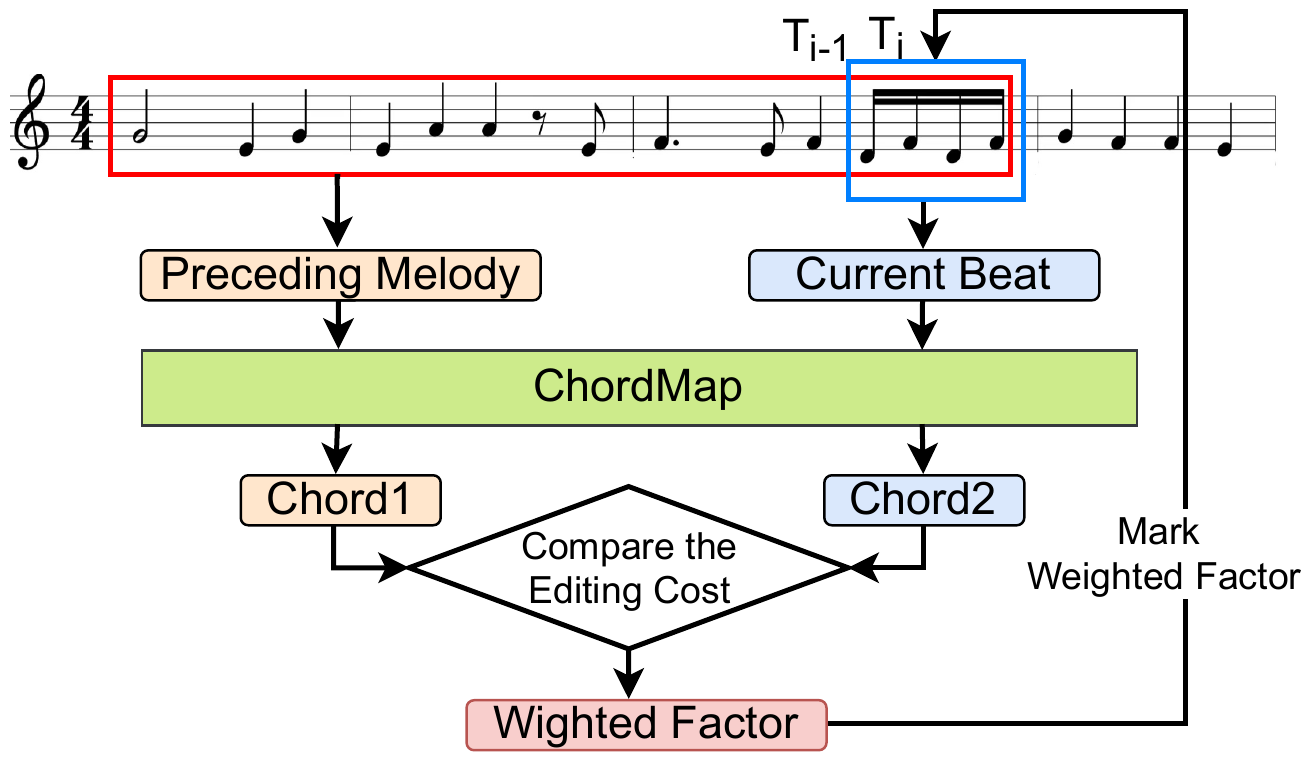}
  \vspace{-0.2cm}
  \caption{The extraction process of weighted factors.}
  \label{WeightedFactor}
  \vspace{-0.2cm}
\end{figure}

Based on Schenker’s theory and the harmonic analysis techniques, we propose the concept of weighted factor(in Fig. \ref{WeightedFactor}). After building a ChordMap ranging from the most basic major and minor triads to polychords, we select a chord from this ChordMap for the current beat with the minimum \textbf{editing cost} using greedy algorithm (in Algorithm \ref{alg:greedy}) as its weighted factor.

\vspace{-0.2cm}
\begin{algorithm}
    \caption{The greedy algorithm in splicing}
    \begin{algorithmic} 
        \State N=[] \qquad \qquad \qquad \# splicing result
        \State N.append(S[-1]) \qquad \qquad \# get current beat
        \State S.pop()
        \State
        \While{True}
            \State $W\_old=w(N)+w(S[-1])+1$
            \State \# calculate cost using only the current beat
            \State $W\_new=w(N+ S[-1])$
            \State \# calculate cost if splicing 
            \If{$W\_new> W\_old$}
            \State \# if the cost increase after splicing,do not splice them
            \State break;
            \Else \qquad \qquad \# otherwise, splice
            \State  Nappend (S[-1])
            \EndIf
            \State S.pop()
        \EndWhile
    \State S.extend(N)  \# put Nback to S
    \end{algorithmic}
    \label{alg:greedy}
\end{algorithm}
\vspace{-0.4cm}

\subsection{Terminal Chord}

\begin{equation}
\begin{aligned}
Terminal\,\, Chord= Chord_{V-I} \cup Chord_{IV-I} \cup \\
Chord_{{V-VI} {sub} {V-I}} \cup Chord_{ \sim -V/VII}  
\end{aligned}
\label{eq:terminal chord}
\end{equation}

A terminal chord marks the termination of a harmony progression at the end of a musical period or an entire piece. The identification of terminal chords is implemented by detecting the harmonic cadences in the chord progression. The common harmonic cadences majorly consist of four types shown in Equation (\ref{eq:terminal chord}). The terminal chords remind Transformer of the chords' positions in the music piece, and play an important role in the texture generation process.

\subsection{Structural Chord}

\begin{equation}
\begin{aligned}
Structural\,\, Chord = {Chord_{non-inverted}} \cap \\
(Chord_{I}\cup Chord_{II}\cup Chord_{IV}\cup Chord_{V})
\label{eq:SC}
\end{aligned}
\end{equation}

Equation (\ref{eq:SC}) shows that structural chords are the $1^{st}$, $2^{nd}$, $4^{th}$ and $5^{th}$chords of the current musical mode and also have to be \textbf{non-inversion}s, which means the chord has the tonic as the lowest note. The structural chords, which are important in the harmonic progression, further provide the model with the long-term musical information.

\section{DATASET}

\subsection{àiSong Dataset}

To enrich the music styles of our dataset, we built the \textbf{àiSong  Dataset}\footnote{https://github.com/CarlWangChina/-iSong-Music-Dataset}  from scratch. We first collected 6,000 guitar scores, where we selected 650 oriental songs that meet our requirements. Furthermore, several music professionals are invited to manually standardize the naming rules and formats of our experimental data. We also split each song into mutually-different sections to reduce data repetition. Segments with different tonalities in the same song will also be listed separately. After 4 months of collection and processing, àiSong Dataset is finally completed, containing 2323 musical pieces.

Our àiSong Dataset is mainly based on the Chinese national pentatonic, which is composed of five positive tones, namely, "Gong(Do), Shang(Re), Jue(Mi), Zhi(Sol) and Yu(La)" and various partial tones. We also transpose the national pentatonic into a natural major with Gong as the tonic.

\subsection{Data Processing of other Datasets}

Apart from our original àiSong Dataset, we also use three open-source datasets (in table \ref{tab:dataset}). These datasets are further standardized following the steps below.

\begin{table}[htbp]
    \caption{The information of datasets}
  \vspace{-0.0cm}
  \label{tab:dataset}
  \begin{tabular}{c|c|c|c}
    \toprule
    Dataset & Musical Pieces & Bars & Duration(hours)\\
    \midrule
    Theorytab & 11270 & 125999 & 105.00\\
    Wikifornia & 4017 & 154643 & 128.87\\
    Nottingham  & 591 & 20575 & 17.15\\
    àiSong &  2323 & 14297 & 11.91\\
    \bottomrule
  \end{tabular}
  \label{tab:statistics}
  \vspace{-0cm}
  \vspace{-0.2cm}
\end{table}

\vspace{-0.2cm}
\subsubsection{Rhythm Screening. }We only reserve the music pieces in 4/4 and 2/4 time for subsequent sampling. To maintain a stable sampling granularity, pieces containing chords shorter than one beat are also deleted.
\vspace{-0.2cm}
\subsubsection{Octave Transposition. }For each piece of music, we calculate the current octave of the melody and accompaniment according to the note distribution. Then by adding to or subtracting several interval differences, we transpose the melody to the $6^{th}$ row of the MIDI standard pitch table and the accompaniment to the $4^{th}$ row.
\vspace{-0.2cm}
\subsubsection{Mode Unification. }Based on the music mode information in the dataset, we convert all major mode music to C major and all minor mode music to A minor. The distinguishment between major mode and minor mode is important because the extraction of our proposed four features is influenced by the mode of current music.

After our processing, the final dataset contains 19,217 musical pieces, including our original àiSong Dataset, which can be directly used for music generation tasks.

\subsection{Music Representation}

We separately sample the melody and accompaniment data in each song into two 1-dimensional time-sequential arrays, which reduces the space complexity and the learning cost of the model.

\begin{figure}[htbp]
  \centering
  \vspace{-0.2cm}
  \includegraphics[width=0.82\linewidth]{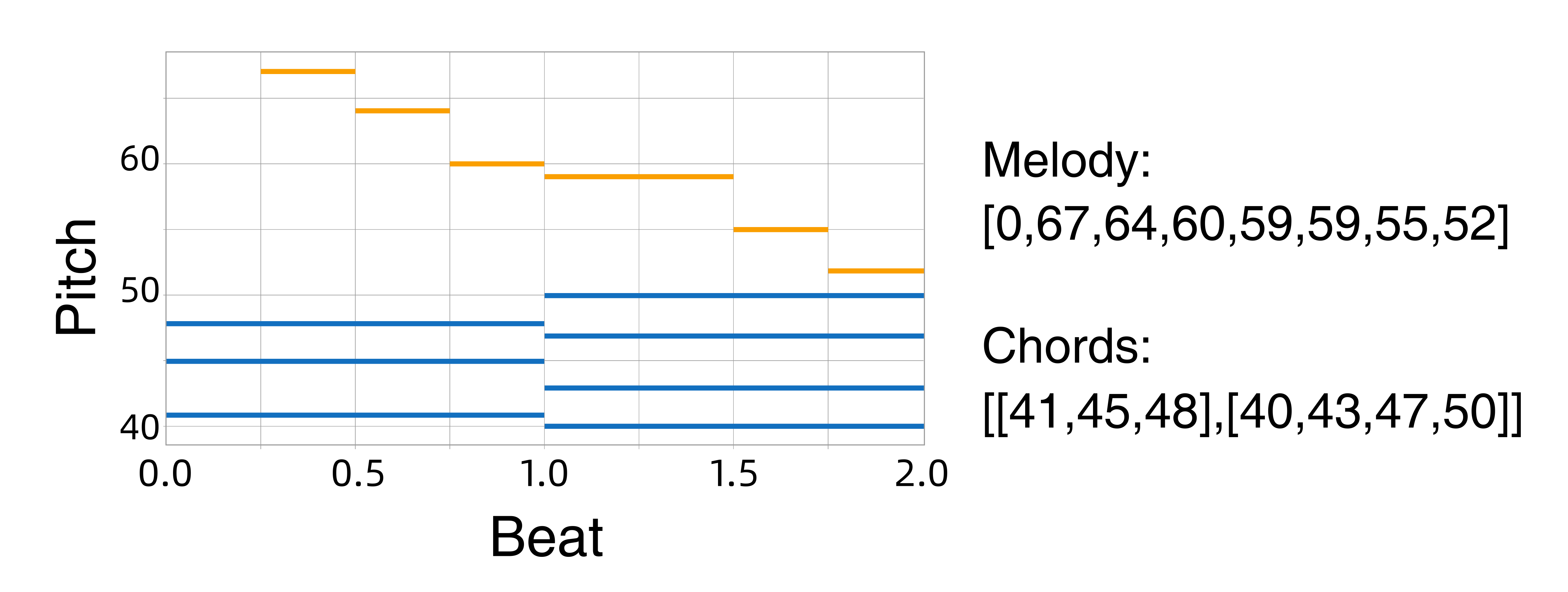}
  \vspace{-0.3cm}
  \caption{  An example of our music representation method. The orange bars (top) represent melody information, and the blue bars (bottom) represent chord information.}
  \label{Representation}
  \vspace{-0.3cm}
\end{figure}

For the \textbf{melody sequence}, the sample rate is every sixteenth note. For \textbf{chord sequence}, we take a sample every quarter note. We use pitch values of notes and chord notes to represent notes and chords, respectively. An example is shown in Fig. \ref{Representation}.

\section{Experiment Setup}
\subsection{Implementation Details}

The detailed architecture of SongDriver is shown in Figure \ref{Overall}. In the experiment, 18367 and 850 clips in the dataset are used for training and testing, respectively.

For the arrangement phase, we employ a Transformer with only one pair of encoder-decoder. The maximum input lengths of the encoder and decoder are set to 4 notes and 3 chords, respectively. We use a single self-attention layer with 8 attention heads. The model hidden size is 256 and the target dimension of the pointwise feedforward net is 2048. In the training, the batch size of one epoch is set to 50 bars, with a learning rate of 0.0001, a dropout rate of 0.1 in each layer and an Adam Optimizer. We train this Transformer for 1000 epochs on a 2080ti GPU with 64GB memory and it takes 33 hours until model convergence. 

For the prediction phase, we train a CRF++0.58 for 20 hours on a cloud server ecs.r6.26xlarge, which has a 104-core CPU with 768G memory. The maximum number of iterations, frequency and cost are set to 35, 3, 4.0, respectively. 

In the real-time cases, the model can only read the inputs ahead of the current time step. To simulate this mechanism in the model training, we adopt a stream input method to step-by-step feed the melody data, instead of giving all musical data to the model at once.

\subsection{Ground Truth}
Previous works usually adopt human-arranged music resources as the ground truth. However, they are not real-time and hence are not comparable to real-time improvisations. Therefore, we invite professional musicians to real-time improvise accompaniments for the 66-minute monophonic melody sample with a BPM(Beat Per Minute) of 80. The improvisations are recorded as the ground truth for the experiment. It is also kept unknown to the musicians before the experiment to ensure the validity of the ground truth.

Though there can be differences among the works by different musicians, the evaluative metrics only focus on the basic evaluation of musicality, instead of emphasizing personal styles. We only intend to compare the accompaniments generated by machines and humans, and find the model with the closest performance to humans.

\begin{table*}[htbp]
\caption{The objective and subjective analyses of model architecture}
\vspace{-0.3cm}
\begin{tabular}{c|c|c|c|c|c|c|c|c|c|c}\toprule
                                \begin{tabular}[c]{@{}c@{}}Objective\\ Metrics\end{tabular}
                                & \begin{tabular}[c]{@{}c@{}}Ground\\ Truth\end{tabular} & \begin{tabular}[c]{@{}c@{}}HMM-\\ LSTM\end{tabular} & \begin{tabular}[c]{@{}c@{}}HMM-\\ Markov\end{tabular} & \begin{tabular}[c]{@{}c@{}}Transformer-\\ Markov\end{tabular} & \begin{tabular}[c]{@{}c@{}}Transformer-\\ LSTM\end{tabular} & \begin{tabular}[c]{@{}c@{}}RNN-\\ Markov\end{tabular} & \begin{tabular}[c]{@{}c@{}}RNN-\\ LSTM\end{tabular} & \begin{tabular}[c]{@{}c@{}}RNN-\\ CRF\end{tabular} & \begin{tabular}[c]{@{}c@{}}HMM-\\ CRF\end{tabular} & \begin{tabular}[c]{@{}c@{}}SongDriver\\ (Transformer-CRF)\end{tabular} \\ \midrule
CTnCTR                          & 0.575                                                  & +0.064                                               & +0.013                                                 & +0.022                                                         & +0.047                                                       & +0.014                                                 & +0.092                                              & -0.034                                             & -0.034                                             & \textbf{-0.010}                                                        \\ 
MCTD                            & 0.478                                                  & +0.018                                               & \textbf{+0.003}                                        & +0.006                                                         & +0.017                                                       & +0.005                                                 & +0.029                                              & -0.009                                             & -0.009                                             & -0.004                                                                 \\ 
HS & 0.208                                                  & -0.028                                              & -0.105                                                & -0.161                                                        & +0.020                                                       & -0.018                                                & +0.147                                              & +0.178                                             & +0.178                                             & \textbf{+0.008}                                                        \\ \bottomrule \end{tabular}
\label{tab:Objective Analysis}
\vspace{+0cm}
\begin{tabular}{c|c|c|c|c|c|c|c|c|c}
\toprule
    \begin{tabular}[c]{@{}c@{}}Subjective\\ Metrics\end{tabular}
    & \begin{tabular}[c]{@{}c@{}}HMM-\\ LSTM\end{tabular} & \begin{tabular}[c]{@{}c@{}}HMM-\\ Markov\end{tabular} & \begin{tabular}[c]{@{}c@{}}Transformer-\\ Markov\end{tabular} & \begin{tabular}[c]{@{}c@{}}Transformer-\\ LSTM\end{tabular} & \begin{tabular}[c]{@{}c@{}}RNN-\\ Markov\end{tabular} & \begin{tabular}[c]{@{}c@{}}RNN-\\ LSTM\end{tabular} & \begin{tabular}[c]{@{}c@{}}RNN-\\ CRF\end{tabular} & \begin{tabular}[c]{@{}c@{}}HMM-\\ CRF\end{tabular} & \begin{tabular}[c]{@{}c@{}}SongDriver\\ (Transformer-CRF)\end{tabular} \\ \midrule
MAH & 2.61                                                & 3.27                                                  & 3.39                                                          & 3.25                                                        & 3.47                                                  & 3.45                                                & 3.10                                               & 3.13                                               & \textbf{3.92}                                                                   \\ 
CPC & 2.90                                                & 3.22                                                  & 3.29                                                          & 3.75                                                        & 3.45                                                  & 3.69                                                & 3.07                                               & 2.94                                               & \textbf{4.08}                                                                   \\ \bottomrule \end{tabular}
\label{tab:Y}
\vspace{-0.3cm}
\end{table*}

\subsection{Objective Evaluation Metrics}
Nine objective metrics are used to evaluate the accompaniments arranged by models. The quality of generated music is not indicated by the absolute scores on these metrics, but by their closeness to the ground truth. 
\textbf{In all the following tables, the scores of models on objective metrics are all their differences from the ground truth.
}

\textbf{1)} \textbf{Average Chord Pitch Interval (CPI)}\cite{yang2020evaluation}, which is the average value of intervals between two consecutive chords;
\textbf{2) Average Chord Inter-Onset Interval (CIOI) \cite{yang2020evaluation},} which finds the average time difference between two consecutive chords;
\textbf{3) Chord Tone to Non-chord Tone Ratio (CTnCTR) \cite{yeh2021automatic},} which calculates the ratio of the number of chord tones to non-chord tones;
\textbf{4) Pitch Consonance Score (PCS) \cite{yeh2021automatic},} which is the tone difference in intervals between the melody and the chord;
\textbf{5) Melody-Chord Tonal Distance (MCTD)} \cite{harte2006detecting}, which calculates the Euclidean distance between the melody vectors and the chord vectors in a 6D linear space; 
\textbf{6) Chord Histogram Entropy(CHE)\cite{yeh2021automatic},} which measures the entropy of a given chord sequence, and reaches maxima when the chord sequence follows a uniform distribution;
\textbf{7) Chord Stability (CS)} is a metric based on Schenker’s theory, which indicates the stability of chord progression. It calculates the difference between the generated chords and the $1^{st}$, $2^{nd}$, $4^{th}$ and $5^{th}$ non-inverted chords of the current music mode, which are important chords in maintaining the stability of the accompaniment;
\textbf{8) Harmonic Structuredness (HS)} calculates the general distribution of harmonic cadences. It reflects the overall structure of arranged accompaniments;
\textbf{9) Weighted Melody-Chord Harmoniousness (WMCH)} Based on Schenker’s music theory, WMCH reflects the difference between the generated chords and the chords composed of the most representative notes in the preceding melody, which indicates the consonance between the original melody and its arranged accompaniment. 








\subsection{Subjective Evaluation Metrics \& Participants}
We employ three different subjective metrics and conduct Mann-Whitney U tests for statistical data comparison.
\textbf{1)MAH:} The harmonious level between the melody and the accompaniment; 
\textbf{2)CPC:} The coherence of the accompaniment’s chord progression;
\textbf{3)MHS:} Melody-Harmonic Synchronization. 

For the subjective evaluations, we randomly select 3 excerpts (beginning, middle, and end) from the 66-minute real-time accompaniments generated by different models. This method can evaluate the changes in music quality with time, which is crucial for assessing the exposure bias problem. In the experiment, we design a questionnaire and invite 17 participants from different backgrounds, including 12 professionals and 5 amateurs. The average time of professional training is 10.57 years. All participants are asked to score the shuffled music excerpts. To simulate the real-time situation, we keep the latency between melody and accompaniment in music samples for subjective evaluation.

\section{Method Analysis}

\subsection{Model Architecture}

To find out the optimal model architecture of SongDriver under the two-phase strategy, we make comparisons among several different combinations of alternative SOTA models, including Transformer, RNN, HMM for the arrangement phase, and CRF, LSTM, Markov for the prediction phase. 

SongDriver (the Transformer-CRF pair) almost outperforms all other model combinations on objective metrics (in Table \ref{tab:Y}). Though narrowly beaten by the HMM-Markov pair on MCTD, SongDriver is overall the best-performed with its shorter processing time.

The scores of SongDriver on subjective metrics (in Table \ref{tab:Y}) are also the highest among all other groups (all p < 0.023), which justifies the best performance of SongDriver again.

\vspace{-0.3cm}
\subsection{Ablation Study}

We use four ablation variants of SongDriver to understand the contribution of each musical feature in real-time accompaniment generation tasks. Each ablation variant corresponds to a different set of our proposed musical features. \textbf{1) SDRW: }SongDriver, removing the weighted factors; \textbf{2) SDRS: }SongDriver, removing the structural chord flags; \textbf{3) SDRT: }SongDriver, removing the terminal chord flags; \textbf{4) SDRN: }SongDriver, removing the weighted notes.

By “removing”, we assign the musical features with their corresponding null values before embedding. 

Both subjective and objective studies (in Table \ref{tab:Ablation}) demonstrate that the model suffers from performance declines without any one of four musical features (with all p < 0.012), which advocates the necessity of introducing four musical features. We suppose this might be caused by the loss of long-term information in real-time conditions.

\begin{table}[h]
\caption{Ablation study results}
\vspace{-0.3cm}
\begin{tabular}{c|c|c|c|c|c|c}
\toprule
\begin{tabular}[c]{@{}c@{}}Obj.\\ Met.\end{tabular}
                                & \begin{tabular}[c]{@{}c@{}}Ground\\ Truth\end{tabular} & SDRW   & SDRS   & SDRT            & SDRN   & \begin{tabular}[c]{@{}c@{}}Song\\ Driver\end{tabular}      \\ \midrule
CIOI                            & 0.247                                                  & +0.341 & +0.365 & +0.356          & +0.355 & \textbf{+0.335} \\ 
CPI                             & 0.288                                                  & +0.398 & +0.420 & +0.417          & +0.414 & \textbf{+0.395} \\ 
HS & 0.208                                                  & -0.015 & +0.055 & \textbf{-0.001} & -0.009 & +0.008          \\ \bottomrule
\end{tabular}
\vspace{+0.4cm}
\begin{tabular}{c|cl|c|c|c|c}
\toprule
\begin{tabular}[c]{@{}c@{}}Subjective\\ Metrics\end{tabular}
    & \multicolumn{2}{c|}{SDRW} & SDRS & SDRT & SDRN & SongDriver \\ \midrule
MAH & \multicolumn{2}{c|}{3.45} & 3.18 & 3.20 & 3.45 & \textbf{3.92}       \\ 
CPC & \multicolumn{2}{c|}{3.63} & 3.43 & 3.33 & 3.53 & \textbf{4.08}       \\ \bottomrule
\end{tabular}
\vspace{-0.3cm}
\vspace{-0.5cm}
\label{tab:Ablation}
\end{table}

\begin{table*}[h]
\caption{The results of the objective evaluation}
\begin{tabular}{c|c|ccc|ccc|c}
\toprule
\multirow{2}{*}{\begin{tabular}[c]{@{}c@{}}Objective\\ Metrics\end{tabular}}
                                &    \multirow{2}{*}{\begin{tabular}[c]{@{}c@{}}Ground\\ Truth\end{tabular}}       & \multicolumn{3}{c|}{Latency Models}                                                                    & \multicolumn{3}{c|}{Bias Models}                                                       &      \multirow{2}{*}{\begin{tabular}[c]{@{}c@{}}Song\\ Driver\end{tabular}}           \\ \cline{3-8}
\multirow{-2}{*}{}              & & \multicolumn{1}{c|}{PSCA-HMM \cite{wallace2019comparing}} & \multicolumn{1}{c|}{PSCA-RNN \cite{wallace2019comparing}} & Music Transformer\cite{huang2018music} & \multicolumn{1}{c|}{Markov-Lin \cite{Lin2021Markov}} & \multicolumn{1}{c|}{Real-CPG \cite{garoufis2020lstm}} & CRF    & \\ 
\midrule
CTnCTR                          & 0.575                          & \multicolumn{1}{c|}{+0.060}           & \multicolumn{1}{c|}{+0.087}           & +0.034                    & \multicolumn{1}{c|}{+0.030}             & \multicolumn{1}{c|}{+0.042}           & -0.037 & \textbf{-0.010}              \\ 
PCS                             & 0.667                          & \multicolumn{1}{c|}{-0.016}          & \multicolumn{1}{c|}{+0.042}           & +0.019                    & \multicolumn{1}{c|}{+0.027}             & \multicolumn{1}{c|}{+0.034}           & -0.016 & \textbf{-0.015}              \\ 
MCTD                            & 0.479                          & \multicolumn{1}{c|}{+0.016}           & \multicolumn{1}{c|}{+0.028}           & +0.013                    & \multicolumn{1}{c|}{+0.011}             & \multicolumn{1}{c|}{+0.013}           & -0.009 & \textbf{-0.004}              \\ 
WMCH                            & 3.296                          & \multicolumn{1}{c|}{\textbf{-0.123}} & \multicolumn{1}{c|}{-0.324}          & -0.445                   & \multicolumn{1}{c|}{-0.256}            & \multicolumn{1}{c|}{-0.732}          & +0.233  & +0.197                        \\ 
CS                              & 1.938                          & \multicolumn{1}{c|}{+0.890}           & \multicolumn{1}{c|}{-0.878}          & +0.929                    & \multicolumn{1}{c|}{-1.776}            & \multicolumn{1}{c|}{+1.046}           & +1.195  & \textbf{+0.874}               \\ 
HS & 0.208                          & \multicolumn{1}{c|}{-0.014}          & \multicolumn{1}{c|}{-0.141}          & +0.028                    & \multicolumn{1}{c|}{-0.204}            & \multicolumn{1}{c|}{+0.192}           & +0.137  & \textbf{+0.008}               \\ \bottomrule
\end{tabular}
\vspace{-0.3cm}
\label{tab:objective evaluation}
\end{table*}

\begin{figure*}[ht] 
	\centering
	\vspace{0.0cm}
    \includegraphics[width=1.03\linewidth]{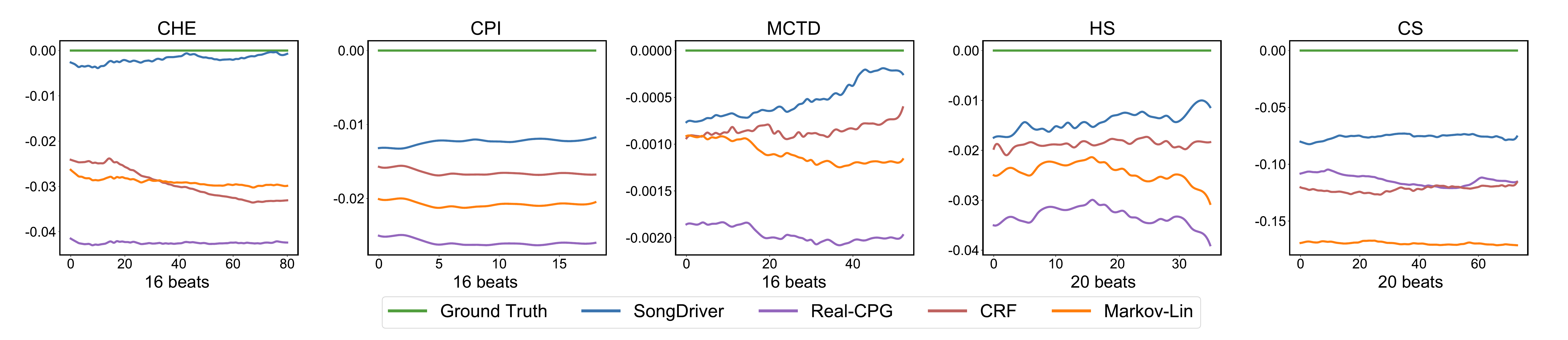}
	\vspace{-0.3cm}
	\caption{The results of the objective evaluation with bias models as the length of the melody sequence increases. }
	\label{PIC:bias models}
	\vspace{-0.1cm}
\end{figure*}

\begin{table*}[ht]
\caption{The results of subjective evaluation}
\begin{tabular}{c|ccc|ccc|c}
\toprule
\multirow{2}{*}{\begin{tabular}[c]{@{}c@{}}Subjective\\ Metrics\end{tabular}} & \multicolumn{3}{c|}{Latency Models}                                               & \multicolumn{3}{c|}{Bias Models}                                                                                  & \multirow{2}{*}{SongDriver}                       \\ \cline{2-7}
                  & \multicolumn{1}{c|}{PSCA-HMM\cite{wallace2019comparing}} & \multicolumn{1}{c|}{PSCA-RNN\cite{wallace2019comparing}} & Music Transformer\cite{huang2018music} & \multicolumn{1}{c|}{Markov-Lin\cite{Lin2021Markov}} & \multicolumn{1}{c|}{Real-CPG\cite{garoufis2020lstm}}                                          & CRF    &                                                   \\ 
\midrule
MAH & \multicolumn{1}{c|}{2.74}     & \multicolumn{1}{c|}{3.45}     & 3.47              & \multicolumn{1}{c|}{3.12 $\rightarrow$}       & \multicolumn{1}{c|}{3.06 $\downarrow$}     & 3.14 $\downarrow$ & \textbf{3.92}  $\rightarrow$     \\ 
CPC & \multicolumn{1}{c|}{2.90}     & \multicolumn{1}{c|}{3.61}     & 3.82              & \multicolumn{1}{c|}{3.12 $\downarrow$}       & \multicolumn{1}{c|}{2.98 $\downarrow$}    & 3.02 $\downarrow$ & \textbf{4.08}  $\rightarrow$     \\ 
MHS & \multicolumn{1}{c|}{3.00}     & \multicolumn{1}{c|}{3.67}     & 3.82              & \multicolumn{1}{c|}{—}          & \multicolumn{1}{c|}{—}        & —    & \textbf{4.14}       \\ \bottomrule
\end{tabular}

\label{tab:subjective evaluation}
\end{table*}

\section{EVALUATION}

\subsection{Objective Evaluation}

We compare the performance of SongDriver with SOTA (state-of-the-art) methods for real-time music accompaniment generation, including 3 latency models and 3 bias models shown in Table \ref{tab:objective evaluation}. Since these SOTA methods are originally evaluated on different metrics, we cannot directly conclude which one should be the best performed. Hence, we list all of them in the contrast group to equally consider their performances.

\subsubsection{Latency Model.}
As Table \ref{tab:objective evaluation} shows, SongDriver has the closest performance to the Ground Truth on almost all metrics. On CTnCTR and MCTD especially, SongDriver can accompany the current melody more harmoniously compared with other latency models, which corroborates that the generated accompaniment of SongDriver is more logically synchronized with the melody. Besides, the CS score also justifies the high harmonic stability of SongDriver generated accompaniments, proving the effectiveness of the two-phase strategy. 
On WMCH, SongDriver is narrowly surpassed by PSCA-HMM. We suppose this might be attributed to HMM's capability to model conditional probabilities, where the dependencies among melodies and chords can be better learned. However, the major problem of HMM-like models lies in its incompetence in capturing contextual dependencies, which may hurt the overall quality of harmonic progression.

\subsubsection{Bias Model. }

Fig. \ref{PIC:bias models} show that SongDriver achieves the most similar effect to the Ground Truth. As the length of melody sequences increases, the curve of SongDriver gradually approaches that of the Ground Truth, while the listed three bias models all exhibit a pattern of performance decline or fluctuation.

\subsection{Subjective Evaluation}

Table \ref{tab:subjective evaluation} shows the mean opinion scores on subjective metrics. The results demonstrate the effectiveness of our approach: SongDriver outperforms other systems on all metrics by a large margin (p < 0.05 for almost all metrics). Though SongDriver exhibits only a slight advantage over Transformer (with a p-value of 0.06), it has achieved a shorter processing time thanks to its two-phase mechanism.

\textbf{Musicality evaluation.} Compared with the listed latency and bias models, SongDriver achieves the highest score on MAH and CPC, the metrics that reflect overall musicality. 

\textbf{Synchronization evaluation.} Participants’ scores on MHS show that the accompaniments generated by the latency models sound more lagging behind the melody. By contrast, the accompaniment generated by SongDriver is more synchronized with melodies, which demonstrates SongDriver’s ability to eliminate the logical latency.

\textbf{Bias evaluation.} According to the participants’ feedback, the musicality of SongDriver-generated accompaniment is more stable, while that of other models almost all declines. Though the score of Markov-Lin on MAH is also stable, it is significantly lower than that of SongDriver. Therefore, SongDriver is still overall the best-performed model.


\section{CONCLUSION}

In this paper, we propose SongDriver, a real-time music accompaniment generation system without logical latency or exposure bias. The innovative combination of the two generation phases compensates for the deficiencies of the latency and bias model. Meanwhile, the four musical features we propose also make up for the loss of long-term sequence information in real-time conditions and can guide SongDriver to generate accompaniments with higher accuracy. In the future, we plan to make SongDriver available for commercial applications, such as developing purchasable hardware products that can provide more tangible experiences to consumers. 

\begin{acks}
This work is partially funded by the Key Project of Natural Science Foundation of Zhejiang Province (No.LZ19F020002), the Key R\&D Program of Zhejiang Province (No.2022C03126) and Project of Key Laboratory of Intelligent Processing Technology for Digital Music (Zhejiang Conservatory of Music), Ministry of Culture and Tourism (No.2022DMKLB001). We also thank Future Design Laboratory of Zhejiang University for the data construction.
\end{acks}

\bibliographystyle{ACM-Reference-Format}
\balance
\bibliography{songdriver}

\newpage
\appendix

\section{TEXTURE PATTERNS}
\label{section:A}

To enrich the auditory effects, we designed multiple texture generation patterns to transform the chords generated during the prediction phase into multi-track accompaniments.

\subsection{Pattern Selection}

First, we need to determine whether the current time step is in the verse or chorus by the rate of pitch change. Next, we need to check if the chords in the preceding 4-16 beats can form a harmonic cadence, which indicates whether the current phrase ends or pauses. Afterward, the texture generation pattern will be selected based on the phrase structure. Where there is a harmonic cadence (switches between phrases or periods), the system plays a piece of decorative accompaniment at the length of one bar. Within a phrase, regular textures will be used in a pre-set order to play accompaniments, where each texture pattern lasts for four bars. We set the switching frequency of textures as per four bars, which is also the length of a basic musical phrase.

\subsection{Process \& Example}

Since most songs often start with verses, the verse piano NO. 1 and NO. 2 are first played in order at the beginning of each song. We then calculate the rate of the pitch change every four bars. If the rate satisfies the conditions of the chorus patterns, then chorus piano NO. 1 (shown in the 5th to 8th bar of the score in Fig. \ref{Texture} in the paper) and chorus guitar, chorus piano NO. 2, and chorus guitar are played in order repeatedly; If the rate meets the conditions of the verse patterns, then verse piano NO. 1 (shown in the 1st bar of the score and the rest 3 bars are the same as it) and verse guitar, verse piano NO. 2 (shown in the 2nd bar of the score and the rest 3 bars are the same as it) and verse guitar, verse piano NO.3 and verse guitar, are played in order repeatedly. 

Additionally, three points need to be noted. First, the lowest pitches of the piano tracks are played by a cello. Second, the score of texture patterns (in the middle of  Fig. \ref{Texture} in the paper) is only an example after combining some chords, because texture patterns cannot directly give expression to any changes in scores. Third, due to the limitation of space, the scores of the guitar, chorus piano NO.2 and verse piano NO.3 are omitted.

\subsection{Texture Representation}

To better represent the texture generation patterns, we use the following textual form to describe each note in the pattern: [The i-th note of the chord from low to high, start time (in beat), duration (in beat), instrument name, intensity]. Here is a simple example: [1, 0, 1, Piano, p5], [2, 1, 1, Piano, p7], [3, 2, 1, Piano, p7], [2, 3, 1, Piano, p7]. This example indicates that the 1st, 2nd, 3rd and 2nd notes of the chord will be played in order on a piano, and the duration of each note is one beat. The intensity of the 1st note is p5 and p7 for the other notes.

For example, the chord generated at the current timestep is suitable for the chorus texture pattern and has reached the fourth bar. Therefore, the notes in the Current Pattern box will be played. These notes can be expressed in a special text format. When generating the final accompaniment, these text formats need to be combined with the currently predicted chord and generate midi files for accompaniment.

\section{MUSICAL FEATURES}
\label{section:B}

This section aims to elaborate on the concepts and extraction methods of the proposed four musical features in the paper.

\subsection{Terminology Explanation}
\textbf{Modes and tonality}: Modes are certain fixed arrangements of tones in an octave, like the major or minor scales. The nature and characteristics of a mode are called tonality. \textbf{Scale degrees of chord}: The note scales in a certain mode are arranged in order, and the scale degree of a chord corresponds to the index of its tonic. For example, the $1^{st}$ scale degree chord ($1^{st}$ chord for short) of C major is a C chord. \textbf{Chord progression}: Two or more chords played in time order. For example, V-I chord progression means to play the $5^{th}$ chord first, and then play the $1^{st}$ chord.

\subsection{Chord Scale Degree}

The extraction of four musical features relies heavily on chord scale degrees, instead of mere chord names. A chord degree can be determined by its name and the current tonality it is under, based on which an algorithm that converts chord names to chord degrees is proposed, shown in Algorithm \ref{alg:chord degree}

\vspace{-0.2cm}
\begin{algorithm}
    \caption{Chord Degree Identification Algorithm}
    \begin{algorithmic} 
        \State Define integer variables: tonality,0,1,2..11 representing C,C\#,D,D\#,E,F,F\#,G,G\#,A,A\#,B,respectively
        \State Define integer variables:mode,0 for minor,1 for major 
        \State Major list=[2,4,5,7,9,11]
        \State Minor list=[2,3,5,7,8,10]
        \State 
        \State Degree=[tonality]
        \If{mode==1}
            \For x in Major list
            \State Degree.append((tonality+x)
            \EndFor
        \EndIf
        \If{mode==0}
            \For x in Minor list
            \State Degree.append((tonality+x)
            \EndFor
        \EndIf    
    \end{algorithmic}
    \label{alg:chord degree}
\end{algorithm}

\subsection{Weighted factor}

\begin{figure}[ht] 
	\centering
	\vspace{0.0cm}
    \includegraphics[width=\linewidth]{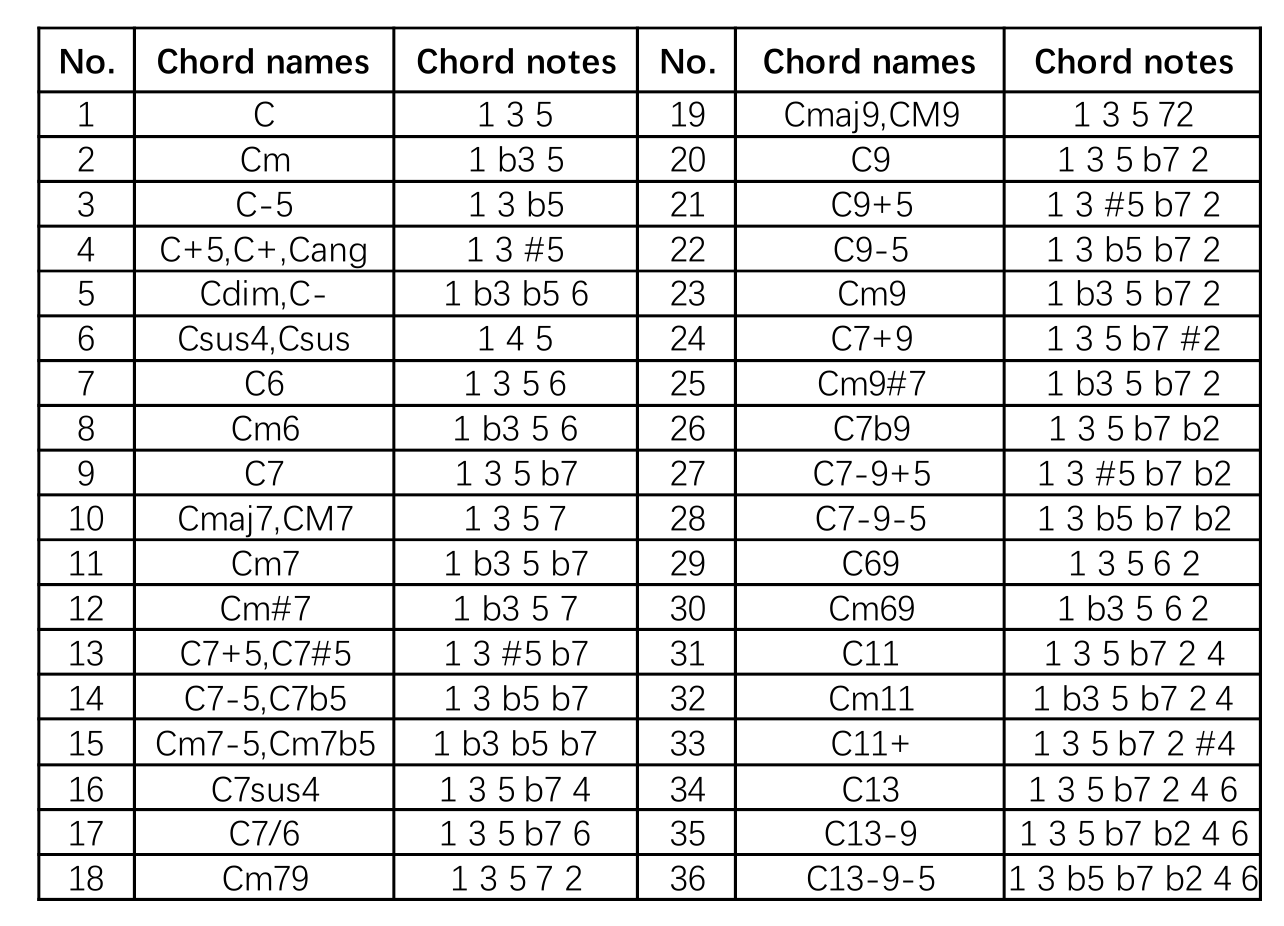}
	\vspace{-0.3cm}
	\caption{One excerpt of the ChordMap.}
	\label{PIC:ChordMap}
	\vspace{-0.3cm}
\end{figure}

\textbf{ChordMap construction.} To find the weighted factors, a chord cross-reference table, namely the ChordMap, is constructed. This Chord-Map ranges from basic major and minor triads to polychords, including 36*12=432 chords. There are 36 chord combinations for each root note, corresponding to 36 different emotional styles. On this basis, the 12 notes in each octave are used as root tones to form 36*12=432 chords. One excerpt of the ChordMap is shown in Fig. \ref{PIC:ChordMap}.

There are two steps in extracting the weighted factors. First, for all the notes in the current beat, the frequency of each note is counted based on this weight allocation: weighted note - 2, other notes - 1, where the first-stage weighted statistics of the current beat melody are calculated. Second, the current beat is concatenated with all the melody fragments in its preceding sequences to form a new melody sequence. The weights of all weighted notes in this melody sequence are set to 2 with those of all other notes to 1. The sum of these weights is the second-stage weighted statistics of the melody sequence. The concatenated sequences are organized in groups of four notes, and each group has the same weighted factor. This concatenation uses the greedy algorithm shown in  Algorithm \ref{alg:greedy} in the paper, where N denotes the final result. The algorithm loops backward from the current note and edits array S following the criterion below: In each step, the N array is examined whether the cost calculation value becomes smaller after it has absorbed S[idx-1]. If the editing cost gets smaller, the greedy strategy continues backwardly. Otherwise, the step is terminated and the current N is the desired fragment. However, the algorithm can terminate at the first loop, when there is only one beat of information in N. To improve the stability, the weight of the current beat is directly assigned with that of the previous beat. Until the next beat is read in, the weighted factors of both beats are then extracted together.

\textbf{Calculating the editing cost.} As shown in Figure \ref{WeightedFactor} in the paper, the editing cost is calculated by comparing the first-stage weighting and the second-stage weighting information with the ChordMap. The editing cost is the sum of minimum intervals between the chord notes from candidate weighted factors and those from the chords in the ChordMap. There are three types of operations: 1) insertion of chord notes, 2) deletion of chord notes, 3) replacement of chord notes.

\subsection{Weighted Note}

The weighted factors are extracted based on the weighted notes. A weighted factor can be regarded as a chord composed of the most representative notes of the corresponding melody, which generalizes the features of the previous tune. Here, we introduce the detailed extraction process of the weighted factor.

In a melody, weighted notes are decisive to the meaning of music, while others can be only for decorative purposes. Since the subsequent melody cannot be foreseen in real-time tasks, a sliding window of a bar’s length is used to simultaneously process the notes in the current beat and its three preceding beats. Then, the statistical information in this sliding window is processed by an algorithm deciding whether the current note is a weighted one or not. The definitions of accents, syncopations, and long notes shown in Equations (\ref{eq:T})(\ref{eq:C})(\ref{eq:L}) in the paper are as follows, all under 4/4-time signature. Accent: The strength of 4/4 time beats in music theory is "strong, weak, next-strong, weak". A note on the strong or the next-strong beat is an accent. Syncopation: A syncopation is a note starting from a weak beat, and lasts until at least half of the following strong or next-strong beat. Long note: Long note refers to the note with the longest duration in each sliding window. If there are multiple longest notes, only the last one in the window is marked as the long note. Weighted notes in paper’s Equation (\ref{eq:WN}) are the notes that are accents but not syncopations, or those that are not accents but are both long notes and syncopations. During the generation process, the weighted notes can mark the importance degree of the current note and optimize the model’s attention distribution.

\subsection{Terminal Chord}

The terminal chords derive from the concept of harmonic cadences, which are used to finish some harmonic progressions at the end of the whole or partial music piece. Harmonic cadences are significant in the chord language, as they often symbolize different musical styles. As shown in Equation (\ref{eq:terminal chord}) in the paper, the identification of the terminal chords is equivalently to detect the harmonic cadences. 
In this paper, harmonic cadences are categorized into four common types: 1) Perfect cadences: the V-I chord progression, which is usually preceded by subordinate chords (II, IV or VI). 2) Plagal cadences: The IV-I chord progression, which highlights the support of the subordinate functional chords to the main chord. 3) Interrupted cadences, where the V7-I progression is replaced by the V7-VI progression. 4) Imperfect cadences: The chord progression from any chord to a V or VII chord.

\subsection{Structural Chord}

Structural chords are the chords that maintain the stability and harmoniousness of generated accompaniments (in Equation (\ref{eq:SC}) in the paper). The chords generated by Transformer must satisfy the following conditions to be identified as structural chords: 1) It must be a non-inverted chord, which means its root note must be its lowest note; 2) Second, the chord must be the $1^{st}$(I), $2^{nd}$(II), $4^{th}$(IV) or $5^{th}$(V) chord of current tonality.
The structural chord extraction process is as follows: First, the input chord is compared with the ChordMap. If the input chord exists in the ChordMap, it is considered as a candidate structural chord. For the first beat, the input chord is the primary chord of the current key. But in the training and generation task, the input chord is the generated chord output by the Transformer model. Then, if the candidate chord is the I, II, IV or V chord in the current key, it is a structural chord.

\end{document}